\documentclass[twocolumn,preprintnumbers,amsmath,amssymb,superscriptaddress]{revtex4}

\usepackage{color}
\usepackage{blindtext}
\usepackage{graphicx}
\usepackage{dcolumn}
\usepackage{bm}
\usepackage{gensymb}
\usepackage{animate}

\def\be{\begin{equation}}
\def\ee{\end{equation}}
\def\bea{\begin{equation*}}
\def\eea{\end{equation*}}
\def\bna{\begin{eqnarray*}}
\def\ena{\end{eqnarray*}}
\def\bn{\begin{eqnarray}}
\def\en{\end{eqnarray}}
\def\bpm{\begin{pmatrix}}
\def\epm{\end{pmatrix}}

\def\be{\begin{equation}}
\def\ee{\end{equation}}
\def\bea{\begin{eqnarray*}}
\def\eea{\end{eqnarray*}}

\newcommand{\bra}[1]{\langle#1|}

\newcommand{\ket}[1]{|#1\rangle}

\newcommand{\braket}[1]{\langle#1\rangle}

\begin{document}

\title{Internal nonlocality in generally dilated Hermiticity}

 \author{Minyi Huang}
 \email{hmyzd2011@126.com}
 \affiliation{Interdisciplinary Center of Quantum Information, State Key Laboratory of Modern Optical Instrumentation, and Zhejiang Province Key Laboratory of Quantum Technology and Device, Department of Physics, Zhejiang University, Hangzhou 310027, China}
 \affiliation{Department of Mathematical Sciences, Zhejiang Sci-Tech University, Hangzhou 310018, PR~China}
\author{Ray-Kuang Lee}
 \email{rklee@ee.nthu.edu.tw}
\affiliation{Institute of Photonics Technologies, National Tsing Hua University, Hsinchu 300, Taiwan}
\affiliation{Department of Physics, National Tsing Hua University, Hsinchu 300, Taiwan}
\affiliation{Center for Quantum Technology, Hsinchu 30013, Taiwan}
\affiliation{Physics Division, National Center for Theoretical Sciences, Taipei 10617, Taiwan}

\begin{abstract}
According to von Neumann, the global Hamiltonian of whole universe must be Hermitian in order to keep the eigenvalues real and to construct a self-consistent quantum theory.
In addition to the open system approach by introducing  environmental degrees of freedom to a small system, a global Hermitian Hamiltonian can also be generated through the dilation from a small Hilbert space.
For example, a local non-Hermitian $\cal PT$-symmetric system can be simulated with a global Hermitian one by the Naimark dilation.
Recently, by introducing local measurements and investigating the correlation functions of outcomes, the internal nonlocality in such dilated Hermitian systems is revealed, but only for a special case with a two-fold structure.
In this paper, we extend such a discussion to the generalized case when the two-fold structure breaks. The internal nonlocality is discussed with different correlation pictures and the corresponding correlation bounds.
Our results provide a device-independent test on the reliability of the simulation in the global Hermiticity.
\end{abstract}

\maketitle

\section{Introduction}
By introducing the environment of infinite free space to an open system, an effective non-Hermitian Hamiltonian can be obtained from a global Hermitian Hamiltonian after eliminating the environmental degrees of freedom~\cite{Feshbach}.
Typically, complex eigenvalues of resonant states are produced due to the coupling between the local subsystem and the macroscopic environment.
However,  parity-time ($\cal PT$) symmetry assumes that the local subsystem can be non-Hermitian but in a parameter regime of real eigenvalues~\cite{bender1998real}.
The symmetry under the combination of time-reversal and parity operations, or more generally an antilinear operation combined with linear operations, can be  generalized to the pseudo-Hermiticity~\cite{mostafazadeh2002Pseudo1,mostafazadeh2002Pseudo2,mostafazadeh2002Pseudo3,mostafazadeh2010pseudo} and anti-$\cal PT$-symmetry~\cite{anti-1, anti-2}, with either real or conjugate pairs of complex eigenvalues.
Lots of theoretical and experimental  applications of $\cal PT$-symmetric systems were found~\cite{El-OL, Makris-PRL, Guo-PRL, Ruter-NP, Chang-NP,Assaw-Nat}, and recently extended to the field of dynamics and  topology ~\cite{ashida2020nonhermitian}.

Similar to the Feshbach formalism dealing with an effective description~\cite{Feshbach}, $\cal PT$-symmetric systems can be viewed as
effective models in the sense of open systems. In 2008, G\"{u}nther and Samsonov showed that a class of unbroken two-dimensional
$\cal PT$-symmetric Hamiltonians can always be dilated to some four-dimensional Hermitian ones ~\cite{gunther2008naimark}.
In fact, by using the dilation techniques, one can simulate any finite dimensional unbroken $\cal PT$-symmetric systems in dilated Hermitian systems~\cite{Lee14, huang2017embedding,PhysRevLett.119.190401, PhysRevLett.123.080404}.
By evolving states under the dilated Hermitian Hamiltonians, it is always possible to simulates the evolution of unbroken $\cal PT$-symmetric Hamiltonians in
 subspaces. On the other hand, for broken $\cal PT$-symmetric systems, their evolutions can also be simulated by utilizing time dependent Hermitian Hamiltonians \cite{Wu878}.

In the simulation of $\cal PT$-symmetric systems, the dilated Hermitian Hamiltonians play an important role, which govern a composite system. By projecting the  dilated Hamiltonians to some subsystems, the effect of $\cal PT$-symmetric Hamiltonians can be realized \cite{huang2017embedding}.
Owing to the non-Hermiticity of $\cal PT$-symmetric systems,
the  dilated Hamiltonians usually bring nonlocal correlations between the subsystems.
Recently, by proposing different correlation pictures, the internal nonlocality of these dilated Hamiltonians were discussed \cite{Huang_2021}.
By evaluating the correlations
with local measurements in three different pictures, the resulting different expectations of the Bell operator
reveal the distinction of the internal nonlocality. Such a result provides the figure
of merit to test the reliability of the simulation, as well as to verify a $\cal PT$-symmetric (sub)system.

However, the known discussions mainly focus on G\"{u}nther and Samsonov's special example,
depending highly on the special form of the dilated Hamiltonian.
In general, such a two-fold structure may not exist in generic dilated Hermitian Hamiltonian. Then how can one have an effective way to verify a more general $\cal PT$-symmetric Hamiltonian when the two-fold structure breaks?
Further more, similar to the device-independent test on the state nonlocality, can we also have detection-loophole-free test on the reliability of the simulation in the global Hermiticity?

In this paper, we propose a generalization of the scenario in Ref.~\cite{Huang_2021}. The correlation pictures are extended in different ways, obtaining the expectations of the Bell operator and their bounds. It is shown that the correlation behaviors are more complex and have new features in the general case.
A direct reflection of this is on the generic Bell operator expectations.
In contrast to the results in Ref. \cite{Huang_2021}, the Bell operator expectations often have some energy shifts in the general case. Moreover, the deviation bounds should be tackled carefully, instead of a simple order relation.
This also leads to the conception of genuine local Hermitian picture in the general case.
Interestingly, the energy shifts and the deviation bounds can also help to distinguish the global Hermitian Hamiltonians.

The remainder of this paper is organized as follows. In Sec. II, we introduce the preliminaries on the related notions
of $\cal PT$-symmetric systems, the concept of dilation,  and the previous known results on the extraction of internal nonlocality~\cite{Huang_2021}.
In Sec. III, we propose different correlation pictures for general Hermitian dilations. The expectations of the Bell operator and their bounds are obtained. Section IV concerns the problem of how to distinguish a dilated Hermitian Hamiltonian. In section V, some discussions are made.
 Finally, we conclude our results in Sec. VI.

\section{Preliminaries}

\subsection{The concept of dilation}
By dilating a time independent $\cal PT$-symmetric Hamiltonian $H$, we mean that one can find some Hermitian operator
$\hat{H}=\begin{bmatrix}H_1&H_2\\H_2^\dag & H_4\end{bmatrix}$ and an invertible Hermitian operator $\tau$ such that for any vector $\psi$,
\be
\begin{bmatrix}
i\psi'\\
i(\tau\psi)'
\end{bmatrix}
=
\begin{bmatrix}
H_1&H_2\\
H_2^\dag&H_4
\end{bmatrix}
\begin{bmatrix}
\psi\\
\tau\psi
\end{bmatrix}
=
\begin{bmatrix}
H\psi\\
\tau H\psi
\end{bmatrix}
.\label{e}
\ee
$\hat{H}$ is called a dilated Hermitian Hamiltonian or a Hermitian dilation of $H$.

Note that for the first component, $i\psi'=H\psi.$
According to the Sch\"{o}dinger equation,
\[\psi(t)=e^{-itH}\psi(0),\] an effective $\cal PT$-symmetric system is realized.

For an unbroken $\cal PT$-symmetric Hamiltonian $H$, one can prove that such an operator $\tau$ always exists and satisfies the following condition, \be
H^\dag(I+\tau^2)=(I+\tau^2) H.\label{condition}
\ee
With $\tau$, one can construct different $\hat{H}$ satisfying Eq. (\ref{e}), among which a typical one is 
\bn
\hat{H}&=&I_2\otimes H_1+i\sigma_y\otimes H_2,\label{Ht1}\\
H_1&=&(H\tau^{-1}+\tau H)(\tau^{-1}+\tau)^{-1},\label{Hl2}\\
H_2&=&(H-\tau H\tau^{-1})(\tau^{-1}+\tau)^{-1}.\label{omega1}
\en
Compared with the general case in Eq. (\ref{e}), the dilated Hamiltonian $\hat{H}$ in Eq. (\ref{Ht1}) only depends on
 $H_1$ and $H_2$. Such a neat and symmetric form (two fold structure) has some interesting properties. For example, one can prove that
 the $\hat{H}$ in Eq. (\ref{Ht1}) has the same eigenvalues as $H$.
For more details of the dilation problem can be referred to Refs. \cite{gunther2008naimark,huang2017embedding,PhysRevLett.119.190401}.

%

\subsection{Two dimensional example}
Let us start with the two-dimensional $\cal PT$-symmetric Hamiltonian ~\cite{gunther2008naimark,bender2007making},
\be H=E_0I_2+s\begin{bmatrix} i\sin\alpha&1\\1&-i\sin\alpha\end{bmatrix}.\label{HG}\ee
The eigenvalues of $H$ are $\lambda_\pm=E_0\pm s\cos\alpha$.
Moreover, there exists an exceptional point when $\sin\alpha=\pm1$ ($\alpha=\pm\frac{\pi}{2}$), in which case the Hamiltonian cannot be diagonalized. When $\alpha \neq \pm\frac{\pi}{2}$,  the Hamiltonian $H$ has real eigenvalues and can be diagonalized. Hence, $\cal
PT$-symmetry is unbroken.
In particular, when $\sin\alpha=0$, i.e. $\alpha=0$ or $\pi$, the Hamiltonian is also Hermitian. In the following, we do not consider the case of broken $\cal PT$-symmetry, since time independent dilation only applies to the case of unbroken $\cal PT$-symmetry. We do not consider the case of Hermitian since it is trivial.

For the $\cal PT$-symmetric Hamiltonian in Eq. (\ref{HG}), a possible way to have the Hermitian dilation $\hat{H}$ is~\cite{gunther2008naimark,huang2017embedding}
\be
\hat{H}=I_2\otimes H_1+i\sigma_y\otimes H_2,\label{Ht}
\ee
where
\bn
&&H_1=E_0I_2+\frac{\omega_0}{2}\cos\alpha \sigma_x,\label{H1'}\\
&&H_2=i\frac{\omega_0}{2}\sin\alpha\sigma_z,\label{omega}\\
&&\omega_0=2s\cos\alpha,
\en
and
\begin{eqnarray}
&&\tau=\frac{1}{\cos\alpha}
\begin{bmatrix}
1 & -i\sin\alpha\\
i\sin\alpha & 1
\end{bmatrix}\label{tau}.
\end{eqnarray}
It can be verified that
the above example is a special case of Eqs. (\ref{Ht1})-(\ref{omega1}). Moreover, $\hat{H}$ has the same eigenvalues as $H$,
with multiplicities of two. According to Eq. (\ref{Ht1}) or Eq. (\ref{Ht}), the dilated Hermitian Hamiltonian $\hat{H}$ is inseparable.
That is, $\hat{H}$ cannot be written as a tensor product of two local operators.
As a consequence, such a global Hamiltonian $\hat{H}$ can bring nonlocal correlations to the subsystems,
which leads to the discussion of nonlocality \cite{Huang_2021}.

\subsection{The internal nonlocality in simulating {\cal PT}-symmetric systems}
A profound approach to discussing the nonlocality is the CHSH (Clauser, Horne, Shimony, and Holt) scenario~\cite{Bellnonlocality,CHSH}. In this scenario, there are two observers Alice and Bob sharing an entangled state, on which they can
perform local measurements. What they want to see is whether the entanglement can bring some nontrivial correlations between the subsystems.
Suppose that Alice can make local measurements $A_1$ and $A_2$, whose outcome is denoted by $a$. Due to the
randomness of the local measurement, the outcome $a$ can take different values, e.g. $a\in \{+1, -1\}$. Similarly, Bob can perform two
measurements $B_1$ and $B_2$ with his outcome $b\in \{+1, -1\}$. A natural way to see the correlations between the outcomes is
 to investigate the expectation value of the product $ab$. For instance, $\braket{A_iB_j}=\sum_{ab} ab P(ab|ij)$ represents the expectation value of $ab$
for given measurements $A_iB_j$, where $P(ab|ij)$ is the joint probability distribution. In particular, one can calculate the expectation of the following Bell operator
\[
S=B_0A_0+B_0A_1+B_1A_0-B_1A_1.\label{bello}
\]
In the classical setting, one does not concern the quantum realization but only a classical (local) description of what Alice and Bob can do. Under this assumption, the probability of Alice and Bob's outcomes do not
depend on each other.
Thus the joint probability distribution admits a product decomposition of Alice and Bob's marginal probability distributions
$p(ab|ij)=\int p(a|i,\nu)p(b|j,\nu)q(\nu)d\nu$, where $\nu$ is some hidden variable with $q(\nu)$ its distribution. In this case, derivations show that $|\braket{S}|\leqslant 2$. However, in the quantum setting, $A_i$ and $B_j$ correspond to different operators and the expectation value is $\braket{S}=Tr(S\rho)$, where $\rho$ is the density operator of the entangled state. Now the bound of $|\braket{S}|$
is $2\sqrt{2}$. The discrepancy between the two bounds shows the difference between the classical (local) and nonlocal correlations, having far reaching influence both theoretically and experimentally. For more details of the standard CHSH inequality, see \cite{Bellnonlocality}.

In the following, we briefly review a CHSH-like discussion on the nonlocal correlations introduced by the dilated Hermitian Hamiltonian of Eq. (\ref{Ht}).
Since now the correlations come from the global Hamiltonian rather than an entangled state, we call it {\it internal nonlocality} to distinguish it from the standard CHSH scenario. Similar to the CHSH's quantum and classical settings giving different Bell operator expectations and bounds, one can also propose different pictures to discuss the internal nonlocality
 \cite{Huang_2021}.
\\

\textbf{The Simulation picture\\}

Suppose Alice and Bob share the dilated Hermitian Hamiltonian $\hat{H}$ in Eq. (\ref{Ht}). Similar to the CHSH scenario, they can make local measurements $A_i$ and $B_j$. Since now we are discussing the correlations introduced by the Hamiltonian, Alice and Bob's ``local measurements'' are actually local states.
Let Alice have the local state $\{\ket{u_+}=u\ket{0}+ v\ket{1}\}$ for $A_0$ and
$\{\ket{u_-}=\overline{v}\ket{0}-\overline{u}\ket{1}\}$ for $A_1$;
while Bob have two local states $\{ \ket{0}\}$ and $\{\ket{1}\}$ for $B_0$ and $B_1$, respectively.
Then the expectations of $B_jA_i$ can be calculated as follows:
\begin{eqnarray}
&&\braket{B_0A_0}=Tr(\ket{0}\bra{0} \otimes \ket{u_+}\bra{u_+})\hat{H},\label{e1}\\
&&\braket{B_1A_0}=Tr(\ket{1}\bra{1} \otimes \ket{u_+}\bra{u_+})\hat{H},\label{e2}\\
&&\braket{B_0A_1}=Tr(\ket{0}\bra{0} \otimes \ket{u_-}\bra{u_-})\hat{H},\label{e3}\\
&&\braket{B_1A_1}=Tr(\ket{1}\bra{1} \otimes \ket{u_-}\bra{u_-})\hat{H}.\label{e4}
\end{eqnarray}
Now, one can further consider the expectation value of the Bell operator:
\bn
&&\braket{B_0A_0}+\braket{B_0A_1}+\braket{B_1A_0}-\braket{B_1A_1}\nonumber \\
&&=2E_0+(\overline{u}v+u\overline{v})\,\omega_0\cos\alpha\label{exs}.
\en
For the last deviation term shown in Eq. (\ref{exs}), we have the following bound
\be
|(\overline{u}v+u\overline{v})\omega_0\cos\alpha|\leqslant|\omega_0\cos\alpha|=|2s\cos^2\alpha|.\label{pers}
\ee\\

\textbf{The Classical picture\\}

The classical picture means that one skips the details of quantum mechanics but only considers a classical description of what Alice and Bob do.
To give such a classical picture,
several key points must be emphasized. Firstly, the classical picture should be consistent with the simulation picture. It requires that Alice has a ``$\cal PT$-symmetric like'' subsystem and the joint measurements of Alice and Bob reveal the characteristics in the global Hamiltonian $\hat{H}$. A natural consequence is to assume that
the measurement results from $A_i$ are just $\lambda_\pm$, namely the eigenvalues of the $\cal PT$-symmetric
Hamiltonian $H$.
Moreover, note that the dilated Hermitian Hamiltonian $\hat{H}$ has the same eigenvalues as the $\cal PT$-symmetric Hamiltonian
$H$ but with a multiplicity of two. Hence the results of $B_i$ should be $1$, such that the correlation functions $B_jA_i$  trivially give the eigenvalues of $\hat{H}$.
Secondly, the ``results'' of Alice and Bob are independent, leading to a classical (nonlocal) correlation. In fact, since Bob's results always give $1$, apparently the two observers' results and the corresponding probability distributions are independent.
Thus,  we do have a classical local picture.

To calculate the expectation of the Bell operator, note that
$\braket{B_jA_i}=\sum_{ab} ab~p(ab|ij)$, where the results $a=\lambda_\pm$ and $b=1$.
Like the standard CHSH scenario,
 one can formally write $p(ab|ij)=\int p(a|i,\mu)p(b|j,\mu)q(\mu)d\mu$, where $\mu$ is a hidden variable with $q(\mu)$ its distribution such that $\int q(\mu)d\mu=1$.
 By changing the variable $d\nu=q(\mu)d\mu$ and denoting $A_i(\nu)=\sum_a ap(a|i, \nu)$, $B_j(\nu)=\sum_b b p(b|j, \nu)$,
 one can see that
  $\braket{B_jA_i}=\int B_j(\nu)A_i(\nu)d\nu$, where $\int 1 d\nu=1$. Unlike the standard CHSH scenario, we only use local states rather than an entangled state, hence the measurements $B_j$ and $A_i$ are completely independent, without an interaction through $\nu$ (or $\mu$). Thus one can assume that $A_i(\nu)$ and $B_j(\nu)$ are constants independent of $\nu$. Moreover, by definition
   $\int A_i(\nu)d\nu=\sum_a a p(a|i)$ and $\int B_j(\nu)d\nu=\sum_b b p(b|j)=1$, hence we see that $A_i(\nu)=\sum_a a p(a|i)$ and $B_j(\nu)=1$. Then
\bn
&&\braket{B_0A_0}+\braket{B_0A_1}+\braket{B_1A_0}-\braket{B_1A_1}\nonumber \\
&&=\int[ B_0(\nu)(A_0+A_1)(\nu)+B_1(\nu)(A_0-A_1)(\nu)]d\nu\nonumber\\
&&=\int[ (A_0+A_1)(\nu)+(A_0-A_1)(\nu)]d\nu\nonumber\\
&&=2E_0+\omega_0\,(p_+-p_-),\label{exc}
\en
where 
$p_\pm=p(\lambda_\pm|0)$
are the probabilities corresponding to the situations when the results of $A_0$ are $\lambda_\pm$.
Moreover,
\be
|\omega_0(p_+-p_-)|\leqslant|\omega_0|=|2s\cos\alpha|.\label{percl}
\ee\\

\textbf{Local Hermitian picture}\\

In this picture, we try to give a description of what Alice and Bob do by some Hermitian Hamiltonians $\hat{H}_l$, which is in a tensor product form of two local Hermitian Hamiltonians. The key concept in this picture is that it should be consistent with 
the simulation. To this end, $\hat{H}_l=I\otimes H_h$, where
$H_h=\lambda_+\ket{s_+}\bra{s_+}+\lambda_-\ket{s_-}\bra{s_-}$ and $\ket{s_\pm}$ are two orthogonal states.
Apparently, due to the form of $\hat{H}_l$, the results of Alice's local measurements
 are $\lambda_\pm$, which is the same as the simulation picture. Moreover, the form of $\hat{H}'$ implies that it will not introduce
nonlocal correlations between the subsystems.


By replacing the Hamiltonian $\hat{H}$ in Eqs. (\ref{e1}-\ref{e4}) with $\hat{H}_l$, the expectation of the Bell operator is
\begin{eqnarray}
&&\braket{B_0A_0}+\braket{B_1A_0}+\braket{B_0A_1}-\braket{B_1A_1}\nonumber\\
&&=2E_0+\omega_0\,(p_+-p_-), \label{exl}
\end{eqnarray}
where $p_\pm=|\braket{u_+|s_\pm}|^2$.

There are two motivations to consider the local Hermitian picture. The first one is to show the difference between a dilated Hermitian Hamiltonian and a global Hamiltonian which will not bring in nonlocal correlations (thus it cannot be used for simulation). Such a global Hermitian Hamiltonian is necessarily a tensor prodct of two local Hermitian Hamiltonians.
The second motivation is to consider a concrete quantum realization of the classical picture.
Since the classical picture is an abstract description of the measurements from the perspective of nonlocal correlations, through the comparison, a quantum realization of the classical picture may give us a deeper understanding on the features of internal nonlocality.  In fact, by comparing Eq. (\ref{exl}) with Eq. (\ref{exs}) and Eq. (\ref{exc}), 
all the expectations in the three pictures contain two terms.
The common term  $2E_0$ is the sum of the two eigenvalues $\lambda_+$ and $\lambda_-$; while the other
one represents a deviation term.
Apparently, the deviations have a relatively simple order relation on their numerical bounds.
The classical and local Hermitian pictures have the same form of deviation term, which usually gives a larger bound than the simulation picture. Such
a result can help to distinguish the dilated Hermitian Hamiltonian \cite{Huang_2021}.

In the above discussions, the special form of the dilated Hamiltonian $\hat{H}$ implicitly plays a key role, especially in the classical and local Hermitian pictures. It renders a reasonable way to correlate Alice and Bob's subsystems, as well as to establish the connections between the
dilated Hamiltonian $\hat{H}$ and the $\cal PT$-symmetric Hamiltonian $H$.
Based on this, one can propose the classical and local Hermitian pictures and compare the corresponding Bell operator expectations in different pictures \cite{Huang_2021}.

\section{The general case}
Now if we are considering the general case, $\hat{H}$ may not have such a special form as Eq. (\ref{Ht}).
Moreover, the dilated Hermitian Hamiltonian $\hat{H}$ may even have eigenvalues different from $H$.
Then can we discuss the internal nonlocality for a general $\hat{H}$?
In the following, we show that this is possible.

Such a generalization is based on the following observation:
If $\hat{H}$ is a Hermitian dilation of a $\cal PT$-symmetric Hamiltonian $H$ (i.e. Eq. (\ref{e}) is valid for $\hat{H}$ and $\tau$), then we have
\be
\begin{bmatrix}
H_1&H_2\\
H_2^\dag&H_4
\end{bmatrix}
\begin{bmatrix}
-\tau\psi\\
\psi
\end{bmatrix}
=
\begin{bmatrix}
-\tau H^\perp\psi\\
H^\perp\psi
\end{bmatrix},\label{taug}
\ee
where
\bn
&&H^\perp=-H_2^\dag\tau+H_4,\label{perp}\\
&&H_2=(H-H_1)\tau^{-1},\label{H2}\\
&&H_4=(\tau H-H_2^\dag)\tau^{-1}.\label{H4}
\en

Utilizing Eqs. (\ref{e}) and (\ref{condition}), one can verify Eqs. (\ref{taug})-(\ref{H4}) through direct calculations
(see Appendix A for details).
In addition, note that there is some freedom to determine the dilated Hamiltonian $\hat{H}$. In fact, Eqs. (\ref{H2}) and (\ref{H4}), show that different Hermitian matrix $H_1$ usually yield different $\hat{H}$.
Eq. (\ref{taug}) also shows that the eigenvalues
of the Hamiltonian $H^\perp$ are just the eigenvalues of the dilated Hermitian Hamiltonian $\hat{H}$.

Eqs. (\ref{e}) and (\ref{taug}) show the effect of the  Hermitian dilation $\hat{H}$ when it is confined to the subsystems. In fact, Eq. (\ref{e}) can be written as
\[
\hat{H}(\ket{0}\ket{\psi}+\ket{1}\ket{\tau\psi})=\ket{0}\ket{H\psi}+\ket{1}\ket{\tau H\psi}.
\]
By post-selecting the ancillary system in state $\ket{0}$, we have
\[
\hat{H}:\ket{0}\ket{\psi}\rightarrow\ket{0}\ket{H\psi}.
\]
Similarly, Eq. (\ref{taug}) can also be written as
\[
\hat{H}(-\ket{0}\ket{\tau\psi}+\ket{1}\ket{\psi})=-\ket{0}\ket{\tau H^\perp\psi}+\ket{1}\ket{H^\perp\psi}.
\]
By post-selecting the ancillary system in state $\ket{1}$, we have
\[
\hat{H}:\ket{1}\ket{\psi}\rightarrow\ket{1}\ket{H^\perp\psi}.
\]
Thus, the effect of $\hat{H}$ can be
represented by two Hamiltonians, one is the $\cal PT$-symmetric Hamiltonian $H$, the other is $H^\perp$.

In particular, it can be verified that if $H_1$ takes the special form in Eq. (\ref{Hl2}), then $H^\perp=H$ and the dilated Hermitian
Hamiltonian $\hat{H}$ reduce to the special case of Eq. (\ref{Ht1}). 
In this special case,
we build up connections between $\hat{H}$ and $H$ to discuss the internal nonlocality. Now for the general case,
both $H$ and $H^\perp$ will be connected with $\hat{H}$. To be more precise,
 one can assume that Alice is either measuring the Hamiltonian $H$ or $H^\perp$ and the
joint measurements of Alice and Bob depict the characteristics of the global Hamiltonian $\hat{H}$. Similar to the discussions in section II, it is possible to reconstruct the different correlation pictures in the general case.\\

\textbf{The simulation picture}\\

In the following, we discuss the correlation pictures for general dilated Hamiltonians of the $\cal PT$-symmetric Hamiltonian $H$
in Eq. (\ref{HG}). For the convenience of discussion, we specify some
notations. Let us denote the special dilated Hamiltonian in (\ref{Ht}) still by $\hat{H}=\begin{bmatrix}H_1&H_2\\H_2^\dag & H_4\end{bmatrix}$, where $H_4=H_1$. Denote a generally dilated Hermitian Hamiltonian by $\hat{H}'=
\begin{bmatrix}H_1'&H_2'\\(H_2')^\dag & H_4'\end{bmatrix}$. When confined to the subspaces, the effect of $\hat{H}'$ can be depicted by $H$ and $(H^\perp)'$.
Denote $\hat{H}''=\hat{H}'-\hat{H}$, then we have $\hat{H}''_i=\hat{H}'_i-\hat{H}_i$. In particular, we denote
\[
H_1''=\hat{H}_1'-\hat{H}_1=
\begin{bmatrix}
a+c&d+ib\\
d-ib&a-c
\end{bmatrix}.
\]

In a general simulation picture, what Alice and Bob do is the same as the special case. They conducts measurements and calculate
 the expectation of the Bell operator, which is given by
 Eqs. (\ref{e1})-(\ref{e4}) with $\hat{H}$ replaced by $\hat{H}'$.
Calculations show that the numerical value of the expectation is (see Appendix B and C for details)
\bn
&&\braket{B_0A_0}+\braket{B_0A_1}+\braket{B_1A_0}-\braket{B_1A_1}\nonumber\\
&=&2E_0+2a+\omega_0''(p_+''-p_-''),\label{exs2'}
\en
where
\be
\omega_0''=2\sqrt{(s\cos^2\alpha+d)^2+\frac{(b+b\sin^2\alpha+2a\sin\alpha)^2}{\cos^4\alpha}+c^2}\label{b1}
\ee
is the difference of the two eigenvalues of $H'_4$,
$p_\pm''=|\braket{u_+|s_\pm''}|^2$ and $\ket{s_\pm''}$ are the eigenstates of $H'_4$.

Comparing Eq. (\ref{exs2'}) with Eq. (\ref{exs}), one can see that now the expectation of the Bell operator has
an energy shift $2a$. For the deviation term, we have
\[
|\omega_0''(p_+''-p_-'')|\leqslant\omega_0''.
\]
\\

 \textbf{The classical picture}\\

The classical picture gives a generic description of what Alice and Bob do,
 regardless of the details of
realization. The key to this picture is that Alice and Bob's results are independent, only with classical (local) correlations.

To have such a classical picture, we consider the following scenario. Suppose Alice can make two measurements $A_0$ and $A_1$.
The results of $A_0$ are the eigenvalues of $H$ and the results of $A_1$ are the eigenvalues of $(H^\perp)'$. However, Alice only knows
that one of $A_i$ outputs the eigenvalues of $H$ and the other outputs the the eigenvalues of $(H^\perp)'$.
Moreover, we assume that Alice makes measurements in some black box. That is, she is unaware of which measurement she has conducted.
Briefly speaking, Alice can only obtain the measurement results but cannot distinguish between $A_0$ and $A_1$.
As for Bob, his results are always $1$. Thus, their
results are independent, implying the correlations are classical (local). Denote
the eigenvalues of $(H^\perp)'$ by $\lambda'_\pm$.
The expectation of Bell operator is (see Appendix B and C for details)
\bn
&&\braket{B_0A_0}+\braket{B_0A_1}+\braket{B_1A_0}-\braket{B_1A_1}\nonumber \\
&=&E_0+E_0'+\frac{1}{2}[\omega_0(p_+-p_-)+\omega_0'(p_+'-p_-')],\label{exc2}
\en
where
\be
E_0'=\frac{1}{2}(\lambda_+'+\lambda_-')=E_0+\frac{2(a+b\sin\alpha)}{\cos^2\alpha}\label{E0'}
\ee
and
\bn
\nonumber&&\omega_0'=\lambda_+'-\lambda_-'\\
\nonumber&=&2\sqrt{(s+\frac{2d}{\cos^2\alpha})^2\cos^2\alpha+\frac{4c^2}{\cos^2\alpha}+\frac{4(b+a\sin\alpha)^2}{\cos^4\alpha}}.\\
\label{b2}
\en
$p'_\pm$ are the probabilities that the measurement results are $\lambda_\pm'$. For the deviation term, we have
\[
|\frac{1}{2}[\omega_0(p_+-p_-)+\omega_0'(p_+'-p_-')]|\leqslant \frac{1}{2}(\omega_0'+|\omega_0|).
\]\\

\textbf{The local Hermitian and genuine local Hermitian picture}\\

There are different approaches to generalizing the local Hermitian picture, which we call local Hermitian and
genuine local Hermitian pictures for the general case.

Still, the local Hermitian picture aims at describing what Alice and Bob do by using local Hermitian Hamiltonians.
As was mentioned, when the ancillary system is post-selected in $\ket{0}$ or $\ket{1}$, the effect of $\hat{H}$ can be represented by
$H$ and $(H')^\perp$ respectively.
Hence to be consistent with the simulation, the global Hamiltonian in the local Hermitian picture is
\be
\hat{H}_l'=\ket{0}\bra{0}\otimes H_h+\ket{1}\bra{1}\otimes (H^\perp)'_h,\label{Hl}
\ee
where $H_h$ and $(H^\perp)'_h$ are Hermitian Hamiltonians having the same eigenvalues as $H$ and $(H^\perp)'$, respectively.
Due to this,
when the ancillary system is in the state $\ket{0}$ or $\ket{1}$,
the measurement results of $\hat{H}'_l$ are the eigenvalues of $H_h$ or $(H^\perp)_h'$, which is similar to $\hat{H}'$.

Suppose that $H_h=\lambda_+\ket{s_+}\bra{s_+}+\lambda_-\ket{s_-}\bra{s_-}$ and $\ket{s_\pm}$ are two orthogonal states.
Similarly, $(H^\perp)'_h=\lambda'_+\ket{s'_+}\bra{s'_+}+\lambda'_-\ket{s'_-}\bra{s'_-}$ and $\ket{s'_\pm}$ are two orthogonal
states.
Replacing $\hat{H}$ with $\hat{H}_l'$ in Eqs. (\ref{e1}) - (\ref{e4}), the expectation of the Bell operator is (see Appendix B and C)
\begin{eqnarray}
&&\braket{B_0A_0}+\braket{B_1A_0}+\braket{B_0A_1}-\braket{B_1A_1}\nonumber\\
&=&2E_0+\omega_0'(p_+'-p_-'),\label{b3}
\end{eqnarray}
where $p'_\pm=|\braket{u_+|s_\pm'}|^2$. Apparently, for the deviation term
\[
|\omega_0'(p_+'-p_-')|\leqslant \omega_0'.
\]

In the special case of Ref.~\cite{Huang_2021}, the local Hermitian picture can be viewed as a quantum realization of the classical picture, yielding the same form of Bell operator expectation. However, Eqs. (\ref{exc2}) and (\ref{b3}) show that they often differ in the general case. In particular, the deviation bounds do not have a simple order relation as that in Ref. \cite{Huang_2021}.

To realize the Bell operator expectation in the classical picture by utilizing local Hermitian
Hamiltonians, we introduce the genuine local Hermitian picture.  The global Hamiltonian is given by
\be
\hat{H}_g'=\frac{1}{2}I\otimes (H_h+ (H^\perp)'_h).\label{lg}
\ee

Replacing $\hat{H}$ with $\hat{H}_g'$ in Eqs. (\ref{e1}-\ref{e4}), the expectation of the Bell operator is (see Appendix B and C)
\bn
\nonumber&&\braket{B_0A_0}+\braket{B_1A_0}+\braket{B_0A_1}-\braket{B_1A_1}\\
&=&E_0+E_0'+\frac{1}{2}[\omega_0(p_+-p_-)+\omega_0'(p_+'-p_-')],\label{exc3}
\en
where $p_\pm=|\braket{u_+|s_\pm}|^2$ and $p'_\pm=|\braket{u_+|s'_\pm}|^2$.

Direct calculations show that the deviation term is smaller than
\be
\sqrt{(\frac{\omega_0}{2})^2+(\frac{\omega'_0}{2})^2+2\frac{\omega_0}{2}\frac{\omega_0'}{2}\cos2\delta},\label{b4}
\ee
where $\delta$ is some parameter related to the angle of $\ket{s_\pm}$ and $\ket{s'_\pm}$.
In particular, when $\frac{\omega_0}{2}\cos2\delta=|\frac{\omega_0}{2}|$ (this is possible when
$\cos 2\delta=1$ or $-1$), the above bound reduces to $\frac{1}{2}(|\omega_0|+\omega_0')$, which is the same as the classical picture.
\section{To distinguish the dilated Hamiltonian $\hat{H}'$}
Suppose we have a set of devices that can produce a dilated Hermitian Hamiltonian and simulate a
$\cal PT$-symmetric system. One may wonder whether the device is reliable, or if it faithfully realizes
the simulation design. Apparently, this question is closely
related to whether the dilated Hermitian Hamiltonian is well prepared.

In particular, it is instructive to distinguish between
the dilated Hermitian Hamiltonian and the local Hermitian. Take the dilated Hamiltonian $\hat{H}$ in Eq. (\ref{Ht}) and
$\hat{H}_l=I\otimes H_h$ in the local Hermitian picture in section II as an example.
Suppose we want to simulate a $\cal PT$-symmetric system, whose effective Hamiltonian is $H$ in Eq. (\ref{HG}). One can
use the device to produce a global system whose Hamiltonian is $\hat{H}$ to realize such a simulation in the subsystem.
Since a $\cal PT$-symmetric system is usually non-Hermitian, the dilated Hermitian Hamiltonian $\hat{H}$ necessarily brings nonlocal correlations to the subsystems.

However, the local Hermitian
Hamiltonian $\hat{H}_l=I\otimes H_h$ cannot produce such correlations, although it has similar properties to the dilated
$\hat{H}$ when local measurements are conducted. Briefly speaking, to see whether the device is reliable, one needs to
distinguish the Hamiltonian it produces from that cannot bring nonlocal correlations. 
 Note that $\hat{H}$ has the same eigenvalues as
  $\hat{H}_l$. Hence one cannot distinguish them only by measurements. However, one can calculate the Bell operator
  expectations.
 Indeed, for the special case of Eq. (\ref{Ht}), it is shown that the classical
 and local Hermitian pictures give larger deviation term than the simulation picture.
  Thus the expectations in different correlation pictures can help to distinguish the dilated Hermitian Hamiltonian and have
  potential applications \cite{Huang_2021}.

  In the general case, one can also discuss how to distinguish
  $\hat{H}'$ from $\hat{H}'_l$. In fact, once the Bell operator expectations differ in the simulation and local Hermitian pictures, one can distinguish
  between $\hat{H}'$ and $\hat{H}'_l$.
  Eqs. (\ref{exs2'}) and (\ref{b3}) show that such a difference may come from two parts, one is an energy shift of $2a$, the
  other is a different bound of the deviation term.

  However, in the general case, the correlation behaviours of the global Hamiltonians are more
  complex and generally one may not distinguish $\hat{H}'$ from $\hat{H}'_l$. In fact, if we take $a=b=c=0$ and $d=\frac{s(\cos^2\alpha-\cos^3\alpha)}{\cos\alpha-2}$, then Eqs. (\ref{b1}) and (\ref{b2}) show that the simulation picture and
  the local Hermitian picture give the same form of Bell operator expectation and deviation bound.
  Thus one cannot distinguish between them.
  Such a result is natural in some sense. In fact,
  $\hat{H}_l'$ in Eq. (\ref{Hl}) is not in a tensor product form. Thus it actually brings nonlocal correlations between the subsystems, just like the Hermitian dilation $\hat{H}$. Hence it is not unexpected that the Bell operator expectations in the two pictures sometimes have the same ranges.

On the other hand, in most cases, such a distinguishment is possible, e.g.
   when $H$ and $(H^\perp)'$ have the same eigenvalues or when $d=0$. In these two cases, one can distinguish between $\hat{H}'$ and $\hat{H}_l'$, by observing either an energy shift or a smaller deviation bound in the   the Bell operator expectation of the simulation picture (see Appendix D for details). This result can be viewed as a natural
   generalization of that in Ref. \cite{Huang_2021}.

    One may also discuss the problem for $\hat{H}'$ and $\hat{H}'_g$. When $\hat{H}'$ and $\hat{H}'_g$ do not have the
    same eigenvalues, one can distinguish them by making measurements directly. When $\hat{H}'$ and $\hat{H}'_g$
    have the same eigenvalues, one can either see an energy shift in the
   the Bell operator expectations or a smaller deviation bound in the simulation picture, which distinguish the two types of
   Hamiltonians.

The above results show that when $\hat{H}'$ and $H$, or equivalently $(H^\perp)'$ and $H$, have the same eigenvalues, the Bell operator expectations have better properties and can help in the task of distinguishing $\hat{H}'$. In fact, since the eigenvalues of
 $H$ are also the eigenvalues of $\hat{H}'$, when $\hat{H}'$ and $H$
 have different eigenvalues, Alice can obtain more than two outcomes. Such a situation is quite different from the standard CHSH scenario. However, when $\hat{H}'$ and $H$ have the same eigenvalues, Alice obtains exactly two outcomes. This is similar to CHSH's discussion, which partly explains why the Bell operator expectations behave better in this case.

\section{Discussions}

Compared to the the special case revealed in Ref.~\cite{Huang_2021}, here,  the physical implications behind the general results are discussed.
A significant difference exists in deriving the classical pictures. To obtain a classical picture for the general case, one needs an extra assumption that Alice cannot distinguish between the measurements $A_i$. Such an assumption is not needed in the special case \cite{Huang_2021}.

However, in the general case, due to the fact that $H$ and $(H^\perp)'$ usually have different eigenvalues, there are four outcomes of measurements, which is quite different from the usual CHSH scenario. If we calculate the usual expectation of the Bell operator, that is, if Alice is aware of the details of his measurements, then she can get an expectation value only related to $H$ but irrelevant to $(H^\perp)'$ (see Appendix C for details). Intuitively, such a biased value is not suitable for investigating the properties of the global Hamiltonian $\hat{H}$.
For this reason, the extra assumption is needed and gives a more reasonable expectation value. Moreover, such an assumption is
implicitly valid for the special Hermitian dilation in Eq. (\ref{Ht}), in which case $(H^\perp)'=H$. Only with the same measurement results,
Alice cannot distinguish between $A_0$ and $A_1$. Hence such an assumption is natural and the classical picture will reduce to the special one in section II.

In the special case of Eq. (\ref{Ht}), the classical and local Hermitian pictures have the same expectation of the Bell operator. However, in the general case, Eqs. (\ref{exc2}), (\ref{b3}) and (\ref{exc3}) show that the Bell operator expectation of the classical picture is different from the local Hermitian picture but the same as the genuine local Hermitian picture. The reason is that the form of $\hat{H}_l'$ is not a tensor product. Although we still use the term ``local
Hermitian picture'' by comparison with the \cite{Huang_2021}, it is actually nonlocal, which cannot be described by a classical picture.
As mentioned, the form of $\hat{H}_l'$ implies that
 when the measurement is $A_0$ (the ancillary system is post-selected in $\ket{0}$),
 the results of Alice are the eigenvalues of $H$.
  When the measurement is $A_1$ (the ancillary system is post-selected in $\ket{1}$),
 the results of Alice are the eigenvalues of $(H^\perp)'$. Since $H$ and $(H^\perp)'$ generally have different eigenvalues,
 Alice can directly distinguish between $A_0$ and $A_1$ by simply reading out the measurement results, contradicting with the assumption that Alice cannot distinguish between the measurements $A_0$ and
$A_1$ in the classical picture. Hence the Bell operator expectation generally differs in the local Hermitian and classical pictures.

To see why
the classical and genuine local Hermitian pictures still have the same expectations, note that if Alice mistaken $A_1$ for $A_0$, then the Hamiltonian should be
\[
\hat{H}_l''=\ket{0}\bra{0}\otimes (H^\perp)'_h+\ket{1}\bra{1}\otimes H_h.\label{1}
\]
Now, since Alice cannot distinguish between $A_i$, the Hamiltonian realizing the Bell operator expectation in the classical picture should be
$\frac{1}{2}(\hat{H}_l'+\hat{H}_l'')=\hat{H}_g'$, which is just Eq. (\ref{lg}). Hence
it is natural that the classical and genuine local Hermitian
pictures have the same expectations.

\section{Conclusion}

In this paper, we investigate the internal nonlocality of generally dilated Hermitian Hamiltonians of $\cal PT$-symmetry. It
is shown that in addition to the $\cal PT$-symmetric Hamiltonian $H$, the effect of a generally dilated Hermitian Hamiltonian can also be characterized by another Hamiltonian $H^\perp$. Based on this observation, the internal nonlocality is revealed in the
general case, even when the two-fold structure of the dilated Hamiltonian in Ref.~\cite{Huang_2021} breaks. Different correlation
pictures are proposed and the Bell operator expectations are obtained.


The results in this paper covers that in Ref.~\cite{Huang_2021}, giving a natural generalization. However, the correlation behaviours are more complex and have new features in the general case.
From the aspect of construction
of correlation pictures, the generic classical picture utilized an assumption which is not needed but implicitly valid for the special case.
A new correlation picture, i.e., the genuine local Hermitian picture, is proposed. The Bell operators often have some energy shifts and
 the deviation bounds are also changed, which do not give a simple order relation as in Ref. \cite{Huang_2021}.
 In particular,
the Bell operator expectation in the local Hermitian picture can coincide with the simulation picture but differs from the classical picture.
It is shown that when the dilated Hermitian and the $\cal PT$-symmetric Hamiltonians have the same eigenvalues, the Bell operator expectations have good properties and can help in the task of distinguishment.
Similar to the device-independent test on the state nonlocality, our results provide a detection-loophole-free test on the reliability of the simulation in a global Hermiticity.

\section*{Acknowledgement}
MH is partially supported by the National Natural Science Foundation of China (11901526), the China Postdoctoral Science Foundation (2020M680074), the Natural Science Foundation of Zhejiang Province (LY22A010010) and the Science Foundation of Zhejiang Sci-Tech University (19062117-Y). RKL is partially supported by the Ministry of Science and Technology, Taiwan under grants (MOST 110-2123-M-007-002, 110-2627-M-008-001),  the International Technology Center Indo-Pacific (ITC IPAC) and Army Research Office, under Contract No. FA5209-21-P-0158, and the Collaborative research program of the Institute for Cosmic Ray Research (ICRR), the University of Tokyo.

\section{Appendix}
\subsection{The properties related to $\hat{H}$ and $H^\perp$}
We show how the Eqs. (\ref{taug})-(\ref{H4}) are obtained. Note that $\tau$ is invertible, Eqs. (\ref{H2}) and (\ref{H4}) are
direct results of Eq. (\ref{e}). To show Eqs. (\ref{H2}) and (\ref{H4}) do give a Hermitian Hamiltonian $\hat{H}$, we have to show $H_4$ is Hermitian. In fact,
$H_4^\dag=H_4$ is equivalent to
\[
(\tau^{-1}) (H^\dag \tau-H_2)=(\tau H-H_2^\dag)\tau^{-1}.
\]
According to Eq. (\ref{H2}), we know $H_2=(H-H_1)\tau^{-1}$. Thus, direct calculations show that the above equation is equivalent to
\[H^\dag(I+\tau^2)=(I+\tau^2) H,\]
which is just Eq. (\ref{condition}). Hence we know that $H_4$ is Hermitian and $\hat{H}$ is a dilated Hermitian Hamiltonian.

To see Eq. (\ref{perp}), note that Eq. (\ref{taug}) is equivalent to the following equations
\begin{eqnarray}
&&-H_1\tau+H_2=-\tau H^\perp,\label{35}\\
&&-H_2^\dag\tau+H_4=H^\perp, \label{36}
\end{eqnarray}
in which Eq. (\ref{36}) is just Eq. (\ref{perp}). Now to show Eq. (\ref{taug}) is valid, we only need to prove
Eq. (\ref{35}). Substituting Eqs. (\ref{H2})-(\ref{H4}) into Eq. (\ref{35}), we find that Eq. (\ref{35}) is also equivalent to
Eq. (\ref{condition}). Thus, Eq. (\ref{taug}) is valid.

It should be noted that Eqs. (\ref{Ht1}) and (\ref{omega1}) can be obtained from Eqs. (\ref{H2}) and (\ref{H4}) when $H_1$ takes the form in Eq. (\ref{Hl2}). In fact, by substituting Eq. (\ref{Hl2}) into Eq. (\ref{H2}), one can obtain $H_2$ in Eq. (\ref{omega1}).

Now to obtain Eq. (\ref{Ht1}), we only need to show $H_4=H_1$, which is equivalent to
\[H_1(\tau+\tau^{-1})\tau=H_4(\tau+\tau^{-1})\tau.\]
According to Eqs. (\ref{H2}) and (\ref{H4}), the above equation can be written as
\[
H_1(\tau+\tau^{-1})\tau=[\tau H-\tau^{-1}(H^\dag-H_1)]\tau^{-1}(\tau+\tau^{-1})\tau.
\]
Now using Eq. (\ref{Hl2}) and the fact $(\tau^{-1}+\tau)\tau=\tau(\tau^{-1}+\tau)$, we see that the above equation reduces to
\[
\tau^{-1}H^\dag(\tau^{-1}+\tau)=\tau H\tau^{-1}+\tau^{-1}H\tau^{-1}.
\]
However, direct calculations show that this equation can be proved by using Eq. (\ref{condition}). Thus we know $H_1=H_4$ and the
dilated Hamiltonian $\hat{H}$ in Eq. (\ref{Ht1}) is indeed a special case when $H_1$ takes the special form in Eq. (\ref{Hl2}).
In fact, one can further prove that $H^\perp=H$ in this special case. To see this, firstly we note that now $H_1=H_4$. Hence it follows from Eq. (\ref{perp}) that
 $H^\perp=-H_2^\dag\tau+H_1$. Thus to show $H^\perp=H$ we only need to prove
 \[
 -H_2^\dag\tau+H_1=H,
 \]
 from which we have
  \[
 (-H_2^\dag\tau+H_1)(\tau^{-1}+\tau)=H(\tau^{-1}+\tau).
 \]
Now according to Eqs. (\ref{Hl2}) and (\ref{H2}), calculations show that the above equation is also equivalent to Eq. (\ref{condition}). Thus we know $H=H^\perp$ in the special case.

To see the eigenvalues of $H^\perp$ are also the eigenvalues of $\hat{H}$, let us assume that $\lambda$ is an eigenvalue of
$H^\perp$ and $\phi$ is an eigenvector. Now Eq. (\ref{taug}) implies that
\[\hat{H}\begin{bmatrix}-\tau\phi\\ \phi \end{bmatrix}=\begin{bmatrix}
-\tau H^\perp\phi\\
H^\perp\phi
\end{bmatrix}=\lambda\begin{bmatrix}
-\tau \phi\\
\phi
\end{bmatrix},
\]
showing that $\lambda$ is an eigenvalue of $\hat{H}$. Similarly, one can show that the eigenvalues of $H$ are also the eigenvalues of
$\hat{H}$. Apparently, if $H^\perp=H$, then $\hat{H}$ has the same eigenvalues as $H$, with multiplicities two.

\subsection{Some calculations related to $\hat{H}'$}
In this part, we show how the expectations in different correlation pictures are obtained.
For the convenience of calculations, denote a general Hermitian dilation Hamiltonian by
$\hat{H}'=
\begin{bmatrix}H_1'&H_2'\\(H_2')^\dag & H_4'\end{bmatrix}$ and the special Hermitian dilation Hamiltonian in Eq. (\ref{Ht}) still by $\hat{H}=\begin{bmatrix}H_1&H_2\\H_2^\dag & H_4\end{bmatrix}$, where $H_4=H_1$. Moreover, assume that $H_i'=H_i+H_i''$ and
\[
H_1''=
\begin{bmatrix}
a+c&d+ib\\
d-ib&a-c
\end{bmatrix}.
\]
Now we see the effect of $H_1''$.
Note that $H_1'=H_1+H_1''$. Then according to Eq. (\ref{H2}),
\begin{eqnarray}
\nonumber H_2'&=&(H-H_1')\tau^{-1}\\
\nonumber &=&(H-H_1)\tau^{-1}-H_1''\tau^{-1}\\
&=&H_2-H_1''\tau^{-1}.\label{H2'}
\end{eqnarray}
According to Eqs. (\ref{H4}) and (\ref{H2'}),
\begin{eqnarray}
\nonumber H_4'&=&(\tau H-H_2^\dag)\tau^{-1}+\tau^{-1}H_1''\tau^{-1}\\
\nonumber &=&H_4+\tau^{-1}H_1''\tau^{-1}\\
&=&H_1+\tau^{-1}H_1''\tau^{-1}.\label{H4'}
\end{eqnarray}
Moreover, Eq. (\ref{taug}) shows that
\bn
\nonumber (H^\perp)'&=&-(H_2')^\dag\tau+H_4'\\
\nonumber &=&-H_2^\dag\tau+H_4+\tau^{-1}H_1''(\tau+\tau^{-1})\\
&=&H+\tau^{-1}H_1''(\tau+\tau^{-1}),
\en
where the last equation holds because $H^\perp=-H_2^\dag\tau+H_4=H$.

Direct calculations show that
\bn
\nonumber&&\tau^{-1}H_1''\tau^{-1}\\
\nonumber&=&\begin{bmatrix}
\frac{a+c+2b\sin \alpha+(a-c)\sin^2\alpha}{\cos^2\alpha}&\frac{d\cos^2\alpha+i(b+b\sin^2\alpha+2a\sin\alpha)}{\cos^2\alpha}\\
\frac{d\cos^2\alpha-i(b+b\sin^2\alpha+2a\sin\alpha)}{\cos^2\alpha}&\frac{a-c+2b\sin \alpha+(a+c)\sin^2\alpha}{\cos^2\alpha}\\
\end{bmatrix},\\ \label{tht}
\en
and
\bn
\nonumber&&\tau^{-1}H_1''(\tau+\tau^{-1})\\
\nonumber&=&\begin{bmatrix}
\frac{2(a+c)+2(b+id)\sin\alpha}{\cos^2\alpha}&\frac{2d+2i(b+a\sin\alpha-c\sin\alpha)}{\cos^2\alpha}\\
\frac{2d-2i(b+a\sin\alpha+c\sin\alpha)}{\cos^2\alpha}&\frac{2(a-c)+2(b-id)\sin\alpha}{\cos^2\alpha}
\end{bmatrix}.\\
\en
Denote
\begin{eqnarray*}
&&A_1=\frac{2(a+b\sin\alpha)}{\cos^2\alpha},\\
&&A_2=\frac{2(b+a\sin\alpha)}{\cos^2\alpha},\\
&&C_1=\frac{2c+2id\sin\alpha}{\cos^2\alpha}+is\sin\alpha,\\
&&C_2=\frac{2d-2ic\sin\alpha}{\cos^2\alpha}+s,
\end{eqnarray*}
then
\[(H^\perp)'=(E_0+A_1)I_2+\begin{bmatrix} C_1& C_2+iA_2 \\ C_2-iA_2 & -C_1\end{bmatrix}.\]
The eigenvalues of $(H^\perp)'$ are
\bn
\nonumber\lambda_+'=E_0+A_1+\sqrt{C_1^2+C_2^2+A_2^2},\label{l+}\\
\nonumber\lambda_-'=E_0+A_1-\sqrt{C_1^2+C_2^2+A_2^2}.\label{l-}
\en
Accordingly, 
\begin{eqnarray*}
&&\nonumber\omega_0'=\lambda_+'-\lambda_-'\\
&&\nonumber=2\sqrt{(s+\frac{2d}{\cos^2\alpha})^2\cos^2\alpha+\frac{4c^2}{\cos^2\alpha}+\frac{4(b+a\sin\alpha)^2}{\cos^4\alpha}},
\end{eqnarray*}
which is just Eq. (\ref{b2}).

Similarly, one can calculate the eigenvalues of $H_4'$ by using Eq. (\ref{H4'}).
Denote
\begin{eqnarray*}
&&A_1'=\frac{a+2b\sin \alpha+a\sin^2\alpha}{\cos^2\alpha},\\
&&A_2'=\frac{b+b\sin^2\alpha+2a\sin\alpha}{\cos^2\alpha},\\
&&C_1'=c,\\
&&C_2'=s\cos^2\alpha+d,
\end{eqnarray*}
then
\[
H_4'=(E_0+A_1')I_2+\begin{bmatrix} C_1'& C_2'+iA_2' \\ C_2'-iA_2' & -C_1'\end{bmatrix}.
\]
Direct calculation show that the two eigenvalues of $H_4'$ are
\bn
\nonumber\lambda_+''=E_0+A_1'+\sqrt{(C_1')^2+(C_2')^2+(A_2')^2},\label{l+}\\
\nonumber\lambda_-''=E_0+A_1'-\sqrt{(C_1')^2+(C_2')^2+(A_2')^2}.\label{l-}
\en
Accordingly, 
\begin{eqnarray*}
&&\nonumber\omega_0''=\lambda_+''-\lambda_-''\\
&&\nonumber=2\sqrt{(s\cos^2\alpha+d)^2+\frac{(b+b\sin^2\alpha+2a\sin\alpha)^2}{\cos^4\alpha}+c^2},
\end{eqnarray*}
which is just Eq. (\ref{b1}).

\subsection{The calculation of different expectations and bounds in the general case}

With the above results, one can calculate the simulation bound.
Note that a general dilated Hermitian Hamiltonian $\hat{H}'$ can be written as
\[
{\hat{H}'}=\ket{0}\bra{0}\otimes H_1'+\ket{1}\bra{1}\otimes H_4'+\ket{0}\bra{1}\otimes H_2'+\ket{1}\bra{0}\otimes (H_2')^\dag.
\]
Denote $H_4'=\lambda_+''\ket{s_+''}\bra{s_+''}+\lambda_-''\ket{s_-''}\bra{s_-''}$,
$\omega_0''=\lambda_+''-\lambda_-''$. By substituting $\hat{H}'$ into Eqs. (\ref{e1})-(\ref{e4}), direct calculations show that
the simulation bound is
\begin{eqnarray*}
&&Tr [\ket{0}\bra{0}\otimes H_1' +\ket{1}\bra{1}\otimes
(\ket{u_+}\bra{u_+}-\ket{u_-}\bra{u_-})H_4']\\
&=&Tr H_1' + \braket{u_+|H_4'|u_+}-\braket{u_-|H_4'|u_-}\\
&=&2E_0+2a+\braket{u_+|H_4'|u_+}-\braket{u_-|H_4'|u_-}\\
&=&2E_0+2a+\omega_0''(p_+''-p_-''),
\end{eqnarray*}
which is just Eq. (\ref{exs2'}).
 The concrete expression of the Bell operator expectation is
\begin{eqnarray*}
\nonumber&&\braket{B_0A_0}+\braket{B_0A_1}+\braket{B_1A_0}-\braket{B_1A_1}\nonumber\\
\nonumber&=&2E_0+2a+(\overline{u}v+u\overline{v})(\omega_0\cos\alpha+2d)\\
\nonumber&+&\overline{u}v\frac{2i(b+b\sin^2\alpha+2a\sin\alpha)}{\cos^2\alpha}\\
&-&u\overline{v}\frac{2i(b+b\sin^2\alpha+2a\sin\alpha)}{\cos^2\alpha}+2c(|u|^2-|v|^2).
\end{eqnarray*}
\\

For the classical picture, the Bell operator expectation is
\begin{eqnarray}
\nonumber&&\braket{B_0A_0}+\braket{B_0A_1}+\braket{B_1A_0}-\braket{B_1A_1}\\
\nonumber&&=\int[ B_0(\nu)(A_0+A_1)(\nu)+B_1(\nu)(A_0-A_1)(\nu)]d\nu\\
&&=\int[ (A_0+A_1)(\nu)+(A_0-A_1)(\nu)]d\nu, \label{41}
\end{eqnarray}
where the last equation hold because the results of $B_i$ are $1$.
Since the results of $A_0$ are the eigenvalues $\lambda_\pm$, then the above equation gives
\be
2E_0+\omega_0(p_+-p_-), \label{exc2'}
\ee
where $p_\pm$ are the probabilities that Alice's results are $\lambda_\pm$.
However, Alice is unaware of the details of the measurements, hence she may mistaken $A_1$ for $A_0$. Thus the
Bell operator expectation will be calculated by changing $A_0$ and $A_1$ in Eq. (\ref{41}). Since the results of $A_1$ are the eigenvalues of $(H^\perp)'$, i.e. $\lambda'_\pm$, hence
the the Bell operator expectation will be
\[
2E_0'+\omega_0'(p'_+-p'_-),
\]
where $E_0'=\frac{1}{2}(\lambda'_++\lambda'_-)$ and $p'_\pm$ are the probabilities that the results are $\lambda'_\pm$.
Now the best Alice can do is to calculate the mean value of the above two results, that is
\[
E_0+E_0'+\frac{1}{2}[\omega_0(p_+-p_-)+\omega_0'(p_+'-p_-')],
\]
which is just Eq. (\ref{exc2}).

We now calculate the Bell operator expectation in the local Hermitian picture. Note that $H_h=\lambda_+\ket{s_+}\bra{s_+}+\lambda_-\ket{s_-}\bra{s_-}$ and $(H^\perp)'_h=\lambda'_+\ket{s'_+}\bra{s'_+}+\lambda'_-\ket{s'_-}\bra{s'_-}$. By replacing $\hat{H}$ with $\hat{H}_l'=\ket{0}\bra{0}\otimes H_h+\ket{1}\bra{1}\otimes (H^\perp)'_h$ in Eqs. (\ref{e1}-\ref{e4}), direct calculations shows
\begin{eqnarray*}
&&\braket{B_0A_0}+\braket{B_1A_0}+\braket{B_0A_1}-\braket{B_1A_1}\nonumber\\
&=&Tr (\ket{0}\bra{0}\otimes H_h+\ket{1}\bra{1}\otimes(\ket{u_+}\bra{u_+}-\ket{u_-}\bra{u_-}) (H^\perp)'_h)\\
&=&2E_0+\omega_0'(p_+'-p_-'),
\end{eqnarray*}
where $p'_\pm=|\braket{u_+|s_\pm'}|^2$. This is just the result of Eq. (\ref{b3}).

Similarly, one can calculate the Bell operator expectation for the genuine local Hermitian picture. Replacing $\hat{H}$ with $\hat{H}_g'=\frac{1}{2}I\otimes (H_h+ (H^\perp)'_h)$ in Eqs. (\ref{e1}-\ref{e4}),
direct calculations show that
\begin{eqnarray*}
&&\braket{B_0A_0}+\braket{B_1A_0}+\braket{B_0A_1}-\braket{B_1A_1}\\
&=&\braket{u_+|H_h|u_+}+\braket{u_+|(H^\perp)'_h|u_+}\\
&=&E_0+E_0'+\frac{1}{2}[\omega_0(p_+-p_-)+\omega_0'(p_+'-p_-')],
\end{eqnarray*}
where $p_\pm=|\braket{u_+|s_\pm}|^2$ and $p'_\pm=|\braket{u_+|s'_\pm}|^2$.  This is just the result of Eq. (\ref{exc3}).\\

We now calculate the bound of Eq. (\ref{b4}) in the genuine local Hermitian picture. Note that such a bound is obtained by altering $\ket{u_\pm}$ but fixing $\ket{s_\pm}$ and $\ket{s'_\pm}$ in Eq. (\ref{exc3}). Since we are now considering two dimensional case, one can parameterize the states in a way similar to the Bloch sphere. Assume that $|\braket{s_\pm|s'_\pm}|=\cos\delta$, where $\delta$ is a parameter characterizing the angle between $\ket{s_\pm}$
and $\ket{s'_\pm}$.
Now without loss of generality, one may assume that $\ket{s_+}=\ket{0}$ and $\ket{s_-}=\ket{1}$. In addition,
\begin{eqnarray*}
&&\ket{u_+}=\begin{bmatrix}
\cos\alpha\\
e^{i\Delta}\sin\alpha
\end{bmatrix}
,
\\
&&\ket{s_+'}=\begin{bmatrix}
\cos\delta\\
e^{i\Delta'}\sin\delta
\end{bmatrix}
,
\ket{s_-'}=
\begin{bmatrix}
-\sin\delta\\
e^{i\Delta'}\cos\delta
\end{bmatrix},
\end{eqnarray*}
where $\Delta$, $\Delta'$ and $\alpha$ are real parameters. Then direct calculations show that
\begin{eqnarray}
\nonumber&&\frac{1}{2}[\omega_0(p_+-p_-)+\omega_0'(p_+'-p_-')]\\
\nonumber&=&\frac{\omega_0}{2}(\cos^2\alpha-\sin^2\alpha)+\frac{\omega_0'}{2}(\cos^2(\alpha+\delta)-\sin^2(\alpha+\delta)\\
\nonumber&+&\sin2\alpha\sin2\delta+\cos(-\Delta+\Delta')\sin2\alpha\sin2\delta)\\
\nonumber&=&\cos2\alpha(\frac{\omega_0}{2}+\frac{\omega_0'}{2}\cos2\delta)+\frac{\omega_0'}{2}\sin2\delta\cos(-\Delta+\Delta')\sin2\alpha\\
\nonumber&\leqslant&\sqrt{(\frac{\omega_0}{2}+\frac{\omega_0'}{2}\cos2\delta)^2+(\frac{\omega_0'}{2}\sin2\delta\cos(-\Delta+\Delta'))^2}\\
&\leqslant&\sqrt{(\frac{\omega_0}{2})^2+(\frac{\omega'_0}{2})^2+2\frac{\omega_0}{2}\frac{\omega_0'}{2}\cos2\delta},\label{bound}
\end{eqnarray}
where the last two inequalities hold due to the Schwartz inequality and the fact $\cos(-\Delta+\Delta')^2\leqslant1$.
When $\frac{\omega_0}{2}\cos2\delta=|\frac{\omega_0}{2}|$, i.e., equivalently
$\cos 2\delta=\pm 1$, Eq. (\ref{bound}) saturates its largest value.
\subsection{To distinguish the Hamiltonian $\hat{H}'$}
When $H$ and $\hat{H}'$, i.e. $(H^\perp)'$ have the same eigenvalues, one can distinguish $\hat{H}'$ from $\hat{H}'_l$ by comparing the Bell operator expectations. Firstly, note that $H$ and $(H^\perp)'$ have the same eigenvalues iff
$\lambda_+'+\lambda_-'=\lambda_++\lambda_-$ and $\lambda_+'-\lambda_-'=|\lambda_+-\lambda_-|$. That is,
\begin{eqnarray*}
E_0'=E_0,~~~\omega_0'=|\omega_0|.
\end{eqnarray*}
Thus, according to Eq. (\ref{E0'}), we have
$
a=-b\sin\alpha.
$
Now if $a\neq 0$, then Eq. (\ref{exs2'}) shows that there is an energy shift $2a$ in the simulation picture. However,  Eq. (\ref{b3}) shows that such a shift does not exist in the local Hermitian picture. Thus by comparing the Bell operator expectations, one can distinguish
$\hat{H}'$ from $\hat{H}'_l$. Now we consider the case $a=0$. Since $a=-b\sin\alpha$, we have $b=0$ ($\sin\alpha=0$ is the trivial case that
$H$ is Hermitian, which is not considered). Now $\omega_0'=|\omega_0|$ implies that
\be
\sqrt{(s+\frac{2d}{\cos^2\alpha})^2\cos^2\alpha+\frac{4c^2}{\cos^2\alpha}}=|s\cos\alpha|.\label{44}
\ee
It follows that
\be
c^2=\frac{\cos^2\alpha}{4}[s^2\cos^2\alpha-(s+\frac{2d}{\cos^2\alpha})^2\cos^2\alpha].\label{c2}
\ee
Note that $c^2\geqslant0$, from Eq. (\ref{c2}) we have
\[
-ds \geqslant \frac{d^2}{\cos^2\alpha}.
\]
Now we calculate $(\omega_0')^2-(\omega_0'')^2$. By substituting Eq. (\ref{c2}) into Eqs. (\ref{b1})
and (\ref{b2}), as well as utilizing the fact $-ds\geqslant d^2/\cos^2\alpha$, we have
\bn
\nonumber&&(\omega_0')^2-(\omega_0'')^2\\
\nonumber&=&4[(s+\frac{2d}{\cos^2\alpha})^2\cos^2\alpha+\frac{4c^2}{\cos^2\alpha}
-(s\cos^2\alpha+d)^2-c^2]\\
\nonumber&=&4s^2\cos^2\alpha-4(s\cos^2\alpha+d)^2-s^2\cos^4\alpha+(s\cos^2\alpha+2d)^2\\
\nonumber&=&4(s^2\cos^2\alpha\sin^2\alpha-ds\cos^2\alpha)\\
\nonumber&\geqslant&4(s^2\cos^2\alpha\sin^2\alpha+d^2)~~~(-ds\geqslant d^2/\cos^2\alpha)\\
\nonumber&>& 0.
\en
Thus we know $\omega_0'=|\omega_0|>\omega_0''$. That is, the local Hermitian picture gives a larger deviation bound than the local Hermitian picture.

To summarize, when $H$ and $(H^\perp)'$ has the same eigenvalues, one can distinguish $\hat{H}'$ from $\hat{H}'_l$ by investigating the Bell operator expectation, there is either an energy shift or a smaller range of
deviation. In particular, the above discussions generalize the results in section II. C, which can be viewed as a special case when  $H=(H^\perp)'$.

Another special case is $d=0$. In this case, Eqs. (\ref{b1}) and (\ref{b2}) show that $\omega_0''\geqslant|\omega_0\cos\alpha|$ and
$\omega_0'\geqslant|\omega_0|$.
By comparing Eqs. (\ref{b1}),
(\ref{b2}), (\ref{b3}) with Eqs. (\ref{pers}), (\ref{exc}), (\ref{exl}), one can see that when $d=0$, the simulation, classical and local Hermitian bounds are usually larger than the special case. Moreover, one can still distinguish $\hat{H}'$ from $\hat{H}'_l$ by
utilizing the Bell operator expectations. In fact, when $a\neq 0$, then there is a shift in the Bell operator expectation for the
simulation picture. Thus one can immediately distinguish $\hat{H}'$ from $\hat{H}'_l$. When $a=0$, by taking $d=0$ in Eqs. (\ref{b1})
and (\ref{b2}), we see that $\omega_0'>\omega_0''$, thus the different bounds of deviation can help distinguish $\hat{H}'$.

One can also discuss the problem for $\hat{H}'$ and $\hat{H}'_g$.
Note that when they have different eigenvalues, one can distinguish between them by measurements.
Hence we only need to consider the situation $\hat{H}'$ and $\hat{H}'_g$ have the same eigenvalues as $H$. Now without loss of generality, one can rewrite the Hamiltonian $\hat{H}'$ in Eq. (\ref{lg}) as $\hat{H}'_g=I\otimes H_h$, where $H_h$ has the same eigenvalues as $H$. Now the Bell
operator expectation of $\hat{H}'_g$ is given by Eq. (\ref{exl}) and the deviation bound is $|\omega_0|$.  Eqs. (\ref{exl}) and (\ref{exs2'})
show that when $a\neq 0$, one can distinguish $\hat{H}'$ from $\hat{H}'_g$ by
the energy shift $2a$. Hence we only need to consider the case $a=0$.
 However, the fact that $\hat{H}'$ and $H$ (or equivalently $(H^\perp)'$ and $H$) have the same eigenvalues also implies that
 $E_0=E_0'$ and $|\omega_0|=\omega_0'$. As shown in the above discussion of distinguishing between
  $\hat{H}'$ and $\hat{H}'_l$, we know $|\omega_0|=\omega_0'>\omega_0''$ in this case. Thus, by comparing the different deviation bounds, one can distinguish $\hat{H}'$ from $\hat{H}'_g$.


%



\end{document}